\documentclass[twocolumn,preprint,nofootinbib]{aastex631}

\usepackage{amsmath}
\usepackage{graphicx}
\usepackage{color}
\usepackage{multirow}
\usepackage{soul}
\DeclareMathAlphabet\mathbfcal{OMS}{cmsy}{b}{n}

\renewcommand\thefootnote{\textcolor{red}{\arabic{footnote}}}

\submitjournal{ApJ}

\begin{document}

\title{Shock-induced partial alignment in geometrically-thick tilted accretion disks around black holes}

\author[0000-0001-5661-7104]{Sajal Gupta}
\affiliation{JILA and Department of Astrophysical and Planetary Sciences, University of Colorado, Boulder, CO 80309, USA}
\correspondingauthor{Sajal Gupta}
\email{Sajal.Gupta@colorado.edu,jason.dexter@colorado.edu}

\author[0000-0003-3903-0373]{Jason Dexter}
\affiliation{JILA and Department of Astrophysical and Planetary Sciences, University of Colorado, Boulder, CO 80309, USA}

\begin{abstract}

We carry out idealized three-dimensional general-relativistic magnetohydrodynamic (GRMHD) simulations of prograde, weakly magnetized, and geometrically thick accretion flows where the gas distribution is misaligned from the black hole spin axis. We evolve the disk for three black hole spins: $a = 0.5, 0.75$, and $0.9375$, and we contrast them with a standard aligned disk simulation with $a = 0.9375$. The tilted disks achieve a warped and twisted steady-state structure, with the outer disk misaligning further away from the black hole and surpassing the initial $24^\circ$ misalignment. However, closer to the black hole, there is evidence of partial alignment, as the inclination angle decreases with radius in this regime. Standing shocks also emerged in proximity to the black hole, roughly at $\sim$ 6 gravitational radii. We show that these shocks act to partially align the inner disk with the black hole spin. The rate of alignment increases with increasing black hole spin magnitude, but in all cases is insufficient to fully align the gas before it accretes. Additionally, we present a toy model of orbit crowding that can predict the location of the shocks in moderate-to-fast rotating black holes, illustrating a potential physical origin for the behavior seen in simulations\textemdash with possible applications in determining the positions of shocks in real misaligned astrophysical systems.
\end{abstract}


\section{Introduction} \label{sec:introduction}
Hot accretion flows have long been of astrophysical interest. These systems, in contrast to their thin-disk counterparts \citep{Shakura}, exhibit significantly reduced luminosity and operate with accretion rates far below the Eddington limit ($\dot{M} \ll \dot{M}_{Edd}$) \citep{Xie_2012, Abramowicz2013}. This makes them suitable to explain the low/hard state of black hole X-ray binaries \citep{Esin_1997, Dexter_2021} and in low-luminosity Active Galactic Nuclei (AGN) including the supermassive Black Hole in our Galactic Centre, Sagittarius A* (Sgr A*) \citep{Yuan2002, Narayan_2008, Feng_2014}. 

Traditionally such systems are assumed to have their angular momentum aligned with the spin of the Black Hole (BH). This presumption holds valid for thin, viscous accretion disks with small tilt angles, allowing viscous torques to gradually bring the inner flow into alignment \citep{Bardeen_Petterson, Liska2018}. However, in the case of large tilt, nonlinear behavior such as the tearing of a disk are observed \citep{Nixon2012,Nealon2015,Liska2021,Kaaz_2023}, and it remains uncertain whether Bardeen-Petterson alignment can occur under these conditions. Alternatively, if the system has a high accretion rate, the spin of the BH may grow and eventually coincide with the angular momentum of the inflowing material \citep{Volonteri_2005}. Furthermore, magneto-spin alignment driven by powerful relativistic jets can also play a crucial role in aligning the strongly magnetized disk with the BH spin \citep{Mckinney_tilted_2013,Chatterjee2023misaligned}. However, the low accretion rates and geometrically thick structure of hot accretion flows such as those observed in Sgr A* and M87 ($\dot{M}$ $\sim 10^{-7}-10^{-9}$ M$_\odot$/yr for Sgr A$^*$ \citep{Dexter2013,White_2020}) make it unlikely that the Bardeen-Petterson effect, cumulative mass accretion, or magneto-spin alignment can induce alignment. Consequently, these systems are more likely to remain misaligned, impacting various physical processes. These include growth rates of supermassive BHs \citep{King2005,Stella1997lense}, their spin measurements, and observable accretion rates. Additionally, the disk misalignment also affects the structure and size of the BH shadow \citep{Dexter2011OBSERVATIONALSO, White_2020, Chatterjee2020} along with the formation and direction of the relativistic jet \citep{King_2018,Liska2018}. For instance, recently \cite{Cui2023} observed that the radio jet in M87 precesses with a $\sim$ 11-year period, attributing this behavior to the misalignment of hot accretion flows.

\cite{Papaloizou_Lin_1995, Nelson_Papaloizou_2000} were among the first to study the evolution of misaligned disks. They developed a linear theory focused on the bending wave regime, and conducted three-dimensional smoothed particle hydrodynamics (SPH) simulations, demonstrating that warps in the low-viscous flows propagate in a wave-like manner rather than diffusing. This causes the disk to remain misaligned even at inner radii, with the disk's plane oscillating as a function of radius \citep{Ivanov_Illarionov_1996,Lubow2002}. Later, \cite{Zhuravlev2011}, while developing an analytical model to describe the structure of twisted disk around slowly rotating Kerr BHs, proposed that these oscillations strongly depend on viscosity in the prograde flows and are notably absent for disks with viscosity parameter, $\alpha$ \citep{Shakura}, between $10^{-3}$ and $10^{-2}$, and effective scale-height, $H/r/\sqrt{a} \sim 10^{-2}$, typically lower than those found in numerical studies of geometrically-thick disks \citep{Fragile_2007,White_2019}. Here, $a \equiv~ cJ_{BH}/GM_{BH}^2$, ranging from 0 to 1, is the dimensionless BH spin parameter. Recently, \cite{White_2019} applied the linear theory to their GRMHD simulations of tilted accretion flows and observed no such oscillations. Instead, the linear theory predicted an exponential increase in disk tilt with decreasing radius, whereas their study showed that disk modestly warps further away from alignment at large radii before a rapid decline in tilt is noticed, highlighting both the limitations of linear bending wave theory and the tendency of the inner flow to partially align. Throughout our study, we refer to this phenomenon as ``partial alignment'', denoting the reduction in inclination angle of the inner disk in its steady-state configuration. Previously, \cite{Sorathia_2013,Sorathia_2014, Hawley_2018} question the relevance of linear theory as they observed that MHD turbulence in misaligned flows disrupts the radial communication of warps, stressing the importance of strong warps driving non-linear effects such as shocks in the system.

\cite{Fragile_2007} carried out simulations of a weakly magnetized hot accretion flow onto a BH of spin $a = 0.9$, and discovered a pair of standing spiral shocks forming close to the BH \citep{Fragile_2008}. The location of shocks found in their study, along with other GRMHD simulations \citep{White_2019,Generozov_2013}, coincides with regions of rapid fall in inclination of the inner flow, raising speculation that the shocks might be responsible for the partial alignment of the inner disk \citep{Mewes2016NumericalRS,Kawaguchi2015}.  Despite this spatial correlation, no conclusive evidence has yet been presented that shocks act to increase the alignment of the inner disk. 

Moreover, the potential mechanism for the formation of shocks in thick, tilted accretion disks is not yet fully understood. Previous studies led by \cite{Fragile_2008} and \cite{Generozov_2013} integrated the fluid trajectories in post-processing and suggested that the convergence of eccentric fluid orbits near the apocenters may be responsible for the shocks formation. However, their analysis were not able to explain the unique geometry of the shocks. It is also unclear how the spatial structure of shocks is affected by the bending and twisting of the disk as well as the spin of the BH. In light of these uncertainties, in this study we attempt to address this speculation of shock-induced partial alignment, and we describe the role of standing shocks in shaping the structure of the disk as well as illustrating a potential mechanism of their origin.

To this end we perform GRMHD simulations. Despite the fact that prior 3D SPH simulations have shown a warp in disk shape \citep{Lubow2002,Nealon2015,Drewes_2021}, none have included a magnetic field in the system. In addition, none of them have observed shocks, which appear to be an important non-axisymmetric feature in thick tilted disks. Previous GRMHD simulations from Fragile and collaborators utilize an artificial viscosity scheme to capture the effects of shocks, whereas the  parameter survey led by \cite{White_2019} was ran for 
insufficient time to attain inflow equilibrium across the inner disk. 

We carry out a series of idealized 3D GRMHD simulations of weakly magnetized, geometrically-thick, tilted accretion disks called standard accretion and normal evolution (SANE) disks, initially making an angle of $24^\circ$ with the BH equatorial plane. We evolve the disks for three different BH spins: a = 0.5, 0.75, and 0.9375, allowing us to study the effect of BH spin on the disk structure. To compare the characteristics of misaligned disks with the conventional picture, we simulated an accretion disk with no initial tilt onto a BH of spin a = 0.9375. We self-consistently evolve an entropy equation for the electrons using the \cite{Ressler_2015} scheme. The latter enables us to calculate the irreversible heating rate on the fly, enabling us to resolve the spatial structure of the shocks. All the cases are evolved with the same initial conditions, thus making the comparison between different models free of this variable. A brief discussion on the simulation setup and initial conditions are given in section \ref{sec:simulation_setup}. We provide our findings in section \ref{sec:results}, and we conclude and discuss our results in section \ref{sec:conclusion}.

\section{Simulation setup} \label{sec:simulation_setup}

We performed idealized 3D GRMHD simulations using the publicly available HARMPI code \footnote{\url{https://github.com/atchekho/harmpi}} (a parallel, 3D version of HARM \citep{Gammie_2003, Noble_2006}) that solves the GRMHD equations for a SANE disk around a rotating BH defined by the Kerr metric with metric signature ($- $,+,+,+). 
We initialize our simulations with a steady-state hydrodynamic equilibrium gas torus \citep{Fishbone1976}, with inner radius and pressure maximum at $r_{in} = 12~r_g$ and $r_{max} = 25~r_g$. Here, $r_g \equiv GM_{BH}/c^2$ denotes one gravitational radii. The torus was seeded with a single loop of poloidal magnetic field, thus facilitating the magneto-rotational instability (MRI) to kick-off. The internal energy and the gas pressure are related by adiabatic index: $P_{gas} = (\Gamma - 1)u_g$, where we chose $\Gamma = 5/3 $. The magnetic field strength is normalized by fixing plasma $\beta$-parameter, $\beta = $max $P_{gas}/$max $P_{b}= 100$, where $P_{b} = b^{\mu} b_{\mu}/2$ is the magnetic pressure. To study the misaligned flows, we rotate the torus with $24^\circ$ about the BH equatorial plane, such that the initial angular momentum of the disk lies in the X-Z plane. We evolve the system in modified spherical-polar Kerr-Schild coordinates ($r,\theta,\phi$), with a grid resolution of 320 $\times$ 256 $\times$ 160 cells to adequately resolve the MRI turbulence. A higher concentration of cells is implemented in the BH equatorial plane to resolve the gas flow and at the pole to resolve the jet at larger radius. Due to the disk's tilt, orbital parcels traverse regions with varying polar resolution. Specifically, the change in polar resolution reaches up to $\Delta \theta \sim 0.05$ radians over the disk extent. In all the cases we studied, the numerical setup covers the region of (0.88 $r_H$, 10$^5~ r_g$) $\times$ (0,$\pi$) $\times$ (0,2$\pi$) in radial, polar and azimuthal directions. Here, $r_H = r_g(1 + \sqrt{1 - a^2})$ is the BH event horizon radius.  The outer radial boundary was extended to $10^5 r_g$ using a superexponential radial coordinate. In our simulations, we use reflective boundary conditions (BCs) in the $\theta$-direction. Unlike transmissive polar BCs, which allow fluid to pass through the boundaries with minimal interference \citep{Liska2018jetprecess}, reflective BCs can induce additional dissipation near the polar axis and potentially enlarge the jet artificially, influencing disk alignment \citep{Mckinney_tilted_2013}. To address these potential issues, we specifically exclude the polar axis region from our averages and remove regions with a magnetization ($\sigma \equiv \frac{b^2}{\rho}$) greater than one.

We utilized \cite{Ressler_2015}'s approach and computed the total heating rate at each time step by measuring the difference between the entropy calculated through energy conservation equations and that from entropy conserving equation. Magnetic and/or kinetic energy dissipated at the grid scale due to the turbulence is recaptured as internal energy, which we use to identify shocks. Our procedure is similar to \cite{White_2019}, with the difference being that we calculated the heating rate at each simulation time-step, which allowed us to incorporate the effects of time-rate change of entropy.

To determine the radial dependence of physical quantities, we follow \cite{Penna2010} averaging approach, and define a density-weighted, time-averaged of quantity X as:
\begin{equation}
    \langle X \rangle_{\rho}(r) = \frac{\int\int\int X~\rho(r,\theta,\phi) dA_{\theta\phi} dt}{\int\int\int \rho(r,\theta,\phi) dA_{\theta\phi} dt}
\end{equation}
where $dA_{\theta\phi} \equiv \sqrt{-g}d\theta d\phi$ is an area-element in spherical polar Kerr-Schild coordinates, and the integral over $dt$ is a time average in the steady-state. The vertically-integrated quantities are computed as 
\begin{equation}
    \langle X \rangle_{\rho,\theta}(r,\phi) = \frac{\int\int X~\rho(r,\theta,\phi) \sqrt{-g}d\theta dt}{\int\int \rho(r,\theta,\phi) \sqrt{-g}d\theta dt}
\end{equation}
We measured the disk thickness using:
\begin{equation}\label{eq:scaleheight}
    H/r = \sqrt{\left\langle \left(\theta^\prime - \theta_0\right)^2 \right\rangle_{\rho,\theta^\prime,\phi^\prime}}
\end{equation}
where $\theta_0 \equiv \pi/2 + \left\langle(\theta^\prime - \pi/2)\right\rangle_{\rho,\theta^\prime,\phi^\prime}$ is the disk midplane location. Here, $(\theta^\prime,\phi^\prime)$ are the angular coordinates aligned with the disk, which are related to the coordinates $(\theta,\phi)$ via Equation \ref{eq:tilted_frame}. Qualitatively, $\theta^\prime = \pi/2$ represents the local midplane of the disk, and $\phi^\prime = 0$ is set to the local precession angle. The notation $\left\langle X \right\rangle_{\rho,\theta^\prime,\phi^\prime}$ denoted a density-weighted, time-averaged value of quantity X in disk aligned (or tilted) coordinates. At later times, the above equation yields $H/r \sim 0.3 - 0.4$ inside $r = 20~r_g$.

\begin{figure*}[t!]
\centering
\includegraphics[width=\textwidth]{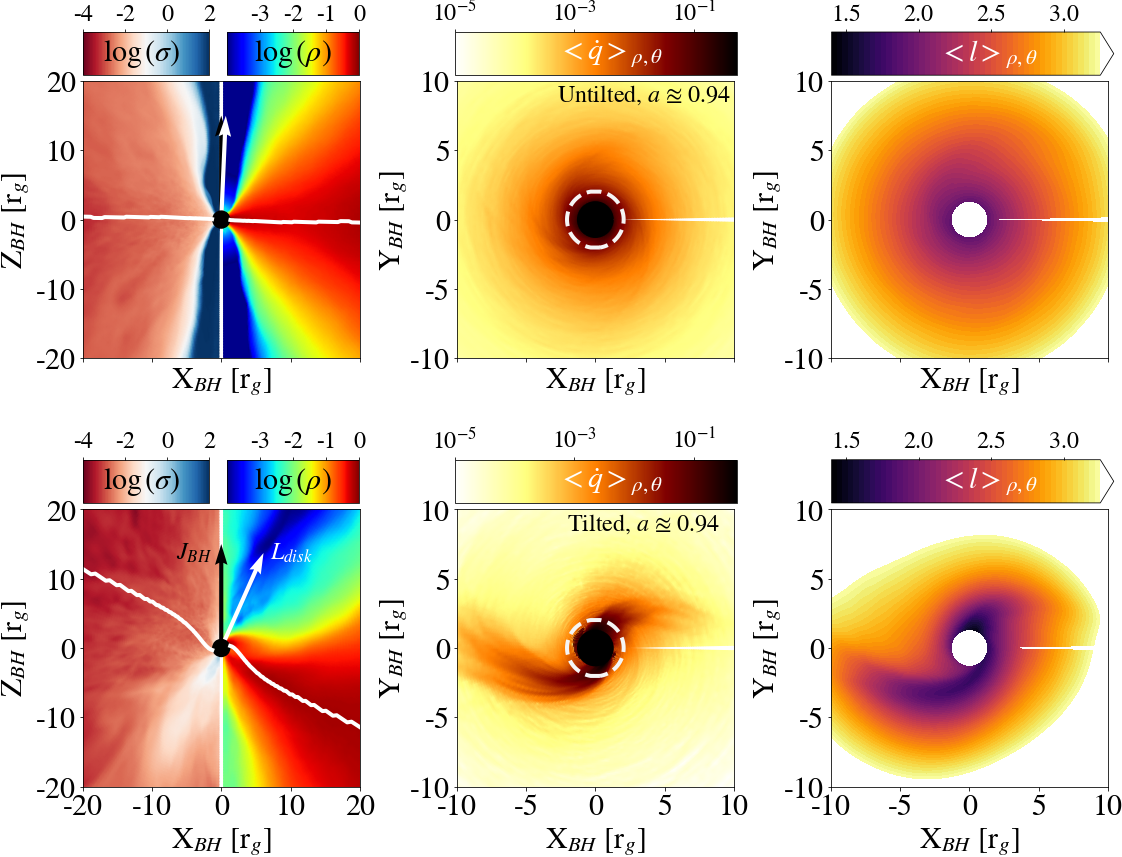}
\caption{\small{Snapshots from the equilibrium states contrasting the properties of a hot, aligned disk (top row) with tilted accretion flows at BH spin, $a \approxeq 0.94$ (bottom row). The color scales are in arbitrary/code units. Bending of the disk is visible in the poloidal slice of the density ($\rho$) and the magnetization ($\sigma$) (leftmost~column), showing the disk midplane (white line), on average, is tilted with respect to equatorial plane of the BH (Z$_{BH}$ = 0). Non-axisymmetric structures in density-weighted, vertically-averaged snapshots of heating rate ($\dot{q}$, middle~column) signify the standing shocks, which extend well beyond the $r_{ISCO}$ (white dashed circular line) and correlates with low specific angular momentum (($l = -u_{\phi}/u_t$, rightmost~column) regions. Here $r_{ISCO}$ denotes the innermost stable circular orbit (ISCO) \textemdash the smallest possible radius for a stable circular orbit.}}
\label{fig:general_characteristics}
\end{figure*}

\vspace{1cm}
\section{Results}
\label{sec:results}
\subsection{General Characteristics: Consequences of an initial tilt}
\label{sec:generalcharacteristics}
The focus of this study is to investigate the steady-state properties of tilted flows. We evolve the simulations until a quasi-steady state is achieved, which we determined by monitoring the density-weighted radial profiles of tilt and twist, as well as the time it takes for standing shocks to fully develop. By these criteria, the tilted disks attain the inflow equilibrium up to $r \sim 20~r_g$ within $t \sim$ $12000~r_g/c$. Tilt reaches equilibrium around $t \sim 10000~r_g/c$, while twist profiles develop from the beginning due to frame-dragging precession. We run the high spin case (a = 0.9375) for $t \sim$ $24000~r_g/c$, to be confident about the steady-state whereas the untilted disk is evolved only for $t \sim$ $10000~r_g/c$, as we notice negligible changes in relevant physical parameters.

\begin{figure*}[t!]
\centering
   \includegraphics[width=1\textwidth]{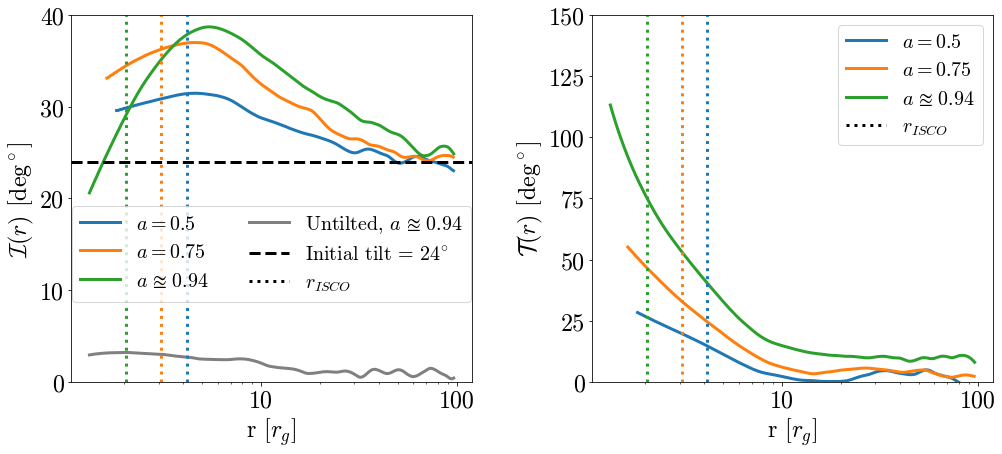}
   \caption{\small{Density-weighted, shell averaged radial profiles of tilt (left) and twist (right) from each simulation (colors). The twist increases as the flow progresses closer to the BH due to frame-dragging precession, which increases with the spin of the BH. The increasing inclination with decreasing radius indicates that the disk warps further away from alignment as it accretes before beginning to align close to the BH (inside $\sim 6~r_g$). This rate of partial-alignment also increases as the spin of the BH increases.}}
   \label{fig:inclination_and_twist}
\end{figure*}

In Figure \ref{fig:general_characteristics}, we showcase and compare misaligned accretion disk profiles with the standard aligned case for spin $a \approxeq 0.94$. The general trend of the disk bending remains similar across other spins, as demonstrated in Appendix \ref{sec:Appendix_A}. The 2D snapshots of different physical quantities for both scenarios are averaged during their respective steady states: from $8000 \leq t \leq 10000~ r_g/c$ for the aligned case (top row) and from $20000 \leq t \leq 22000~ r_g/c$ for the misaligned case (bottom row).  Since both simulations achieved steady states, averaging at different times does not represent distinct evolutionary stages.

The left column shows the poloidal slice of the rest-mass density ($\rho$) and the magnetization ($\sigma \equiv \frac{b^2}{\rho}$) for aligned and misaligned flows. Comparing these profiles reveals a pronounced disk bending in the tilted flow, indicating persistent misalignment at the steady-state. Perpendicular to the disk midplane (white arrow) is its angular momentum vector ($L_{disk}$), pointing toward the low-density, high-magnetization region called the funnel region. This suggests that jets may emit in the direction of disk angular momentum, and may precess with the disk as remarked by \cite{Liska2018jetprecess}. Near the BH, the tilted flow shows two separate density arms, which are the regions of high compression formed due to standing shocks \citep{Fragile_2008, Generozov_2013}. 

To identify standing shocks and their effects, we plot density-weighted, vertically averaged snapshots of the total heating rate ($ \equiv \dot{q}$), and the specific angular momentum ($l \equiv -u_{\phi}/u_t$) in the central and right columns of Figure \ref{fig:general_characteristics}. In the aligned case (top row), both $\langle\dot{q}\rangle_{\rho,\theta}$ and $\langle l \rangle_{\rho,\theta}$ are axisymmetric as expected. However, in the tilted flow (bottom row), the structure completely changes, and we find significant non-axisymmetric features in both quantities. Furthermore, the deviations from axisymmetry increase with BH spin, as illustrated in Figure \ref{fig:Appendix:general_characteristics}. 

The m = 2 azimuthal structure in heating rate ($\dot{q}$) plots (cf. figure \ref{fig:general_characteristics}) signifies the pair of standing, spiral shocks. Here the m = 2 mode signifies the Fourier decomposition of the azimuthal structure in the heating rate. These shocks are localized, stronger, and have a more well-defined structure at higher spins. Comparing the heating rate and specific angular momentum in the tilted flow reveals that regions of depleted angular momentum correspond to the standing shocks, suggesting that they may enhance the transport of angular momentum. The geometry of the shocks in the disk-aligned frame is illustrated in Appendix \ref{sec:Appendix_E}.
\subsection{Inclination and twist}

\begin{figure*}[t!]
\centering
\includegraphics[width=\textwidth]{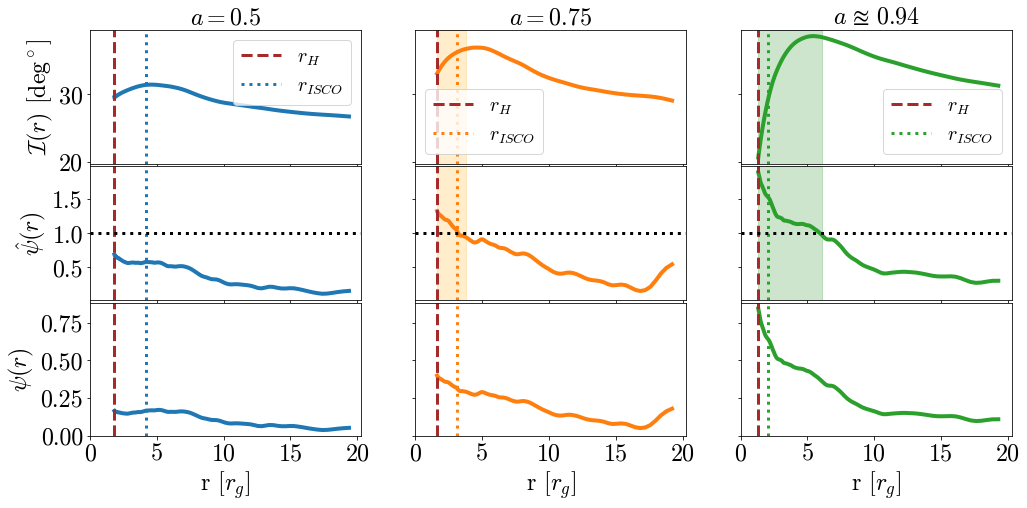}
\caption{\small{Density-weighted, shell averaged radial profiles of tilt (top row), dimensionless warp ($\hat{\psi}$) (middle row) and warp amplitude (bottom row) for tilted simulations (colors).  At small radii (shaded areas), the simulations exhibit increased warping ($\hat{\psi}(r) \geq 1$), indicating the presence of non-linear effects, which coincide with regions of decreasing tilt (top row).}}
\label{fig:warp}
\end{figure*}

\label{sec:inclinationandtwist}
We measure the tilt as a function of disk radius by defining inclination as:
\begin{equation}
    \mathcal{I}(r) \equiv \cos^{-1}\left(\frac{\mathbf{J_{BH}}\cdot\mathbf{L_{disk}(r)}}{|\mathbf{J_{BH}}||\mathbf{L_{disk}(r)}|}\right)
\end{equation}
where $\mathbf{J_{BH}}$ = $a~\hat{z}$ is the angular momentum of BH in natural units. The angular momentum of the disk: $\mathbf{L_{disk}}$ is defined as:
\begin{equation}
    (\mathbf{L_{disk}})_{\hat{k}} = \epsilon_{ijk} r^{\hat{j}}~(T^t~_{\hat{k}})_{MA}
\end{equation}
where $\epsilon_{ijk}$ is the three-dimensional anti-symmetric Levi-Civeta symbol, and the hats indicate the cartesian coordinates related to spherical coordinates in the conventional way. We call $(L_x, L_y, L_z) \equiv$  $(\mathbf{L_{disk}}_{(\hat{x})},\mathbf{L_{disk}}_{(\hat{y})},\mathbf{L_{disk}}_{(\hat{z})})$.  Using the above information, we can write density-weighted, average inclination as:
\begin{equation}
    \mathcal{I}(r) = \cos^{-1}\left(\frac{\langle L_z \rangle_{\rho}}{\sqrt{\langle L_x \rangle_{\rho}^2 + \langle L_y \rangle_{\rho}^2 + \langle L_z \rangle_{\rho}^2}}\right)
\end{equation}
and the twist as:
\begin{equation}
    \mathcal{T}(r) = \tan^{-1}\left(\frac{\langle L_y \rangle_{\rho}}{\langle L_x \rangle_{\rho}}\right)
\end{equation}

The inclination $\mathcal{I}(r)$ is the measure of how elevated the disk midplane is from the BH equatorial plane, whereas the twist $\mathcal{T}(r)$ is the measure of the precession of the disk's angular momentum about the spin axis of BH. If the inclination and/or twist is a function of radius, the disk acquires a warped structure and is referred to as a warped disk as mentioned in section \ref{sec:introduction}.

Figure \ref{fig:inclination_and_twist} shows the radial, steady-state profiles of averaged inclination ($\mathcal{I}(r)$) [left panel] and twist ($\mathcal{T}(r)$) [right panel]. The radial profiles of angles are averaged in their steady-state, particularly, $8000 \leq t \leq 10000$ for the untilted case, $12000 \leq t \leq 14000$ for $a = 0.5$, and $0.75$, and  $20000 \leq t \leq 22000$ for the high spin case ($a \approxeq 0.94$). To get a sense of the accretion flow, it is best to read the plot from right to left. The twist sets up very early in the simulation due to frame-dragging precession and later relaxes to a smaller value due to the transport of angular momentum \citep{White_2019}. The stronger twist in the high spin case is because frame-dragging precession is proportional to the spin of the BH \citep{Lense-Thirring}. Overall, the shape of our twist profile indicates that the disk becomes extremely twisted as one approaches the BH.  

In the left panel of Figure \ref{fig:inclination_and_twist}, a tiny tilt of approximately 3$^\circ$ (grey line) is noticeable in the aligned case. Ideally, we expect a flat line of 0$^\circ$. This discrepancy arises due to numerical errors associated with taking the projections of
the angular momentum of the disk on the BH equatorial plane, as the out-of-plane angular momentum projections, namely $L_x$ and $L_y$, are not conserved. Nevertheless, the associated error is small, suggesting that our routine for determining the inclination is accurate to within about $\mathcal{O}(3^\circ$). Our definition also yields the initial tilt and twist far away from the BH, as we expect the tilt and twist to asymptote to an initial value of 24$^{\circ}$ and 0$^{\circ}$, which we can see is true for all cases.

Analyzing the tilted profiles reveals that inclination increases as one approaches the BH, implying that the disk progressively bends further away from the BH's equatorial plane. Given the simulations' duration, which exceeds the inflow timescale, we conclude that the disk's warping is a stable characteristic, associated with bending waves. However, the predicted high-frequency oscillations from linear theory are virtually absent in our tilt profile.  This could be due to the small viscosity in our flows, which may suppress the large oscillations and mitigate the radial communication of warps \citep{Zhuravlev2011}. 

The rapid decline in inclination, especially at very small radii ($r \leq$ $6~r_g$), causes the disk to partially align with the BH. This descent occurs at a significantly faster rate when the BH has a high spin, around $a \approxeq 0.94$, indicating a direct influence of the BH's spin on the inner disk's physical structure. In our observations, we also confirm the presence of shocks in the disk. However, before quantifying their impact on the tilt profile, if any, it is essential to determine their radial extent.

We achieve this by identifying the region where the dimensionless warp parameter, represented as $\hat{\psi} = \psi/(H/r)$, exceeds 1. Here $\psi(r) \equiv \left|\frac{r d \mathbf{l(r)}}{d r}\right|$ signifies the measure of local warp and $\mathbf{l}(r)$ = $\mathbf{L_{disk}(r)}/|\mathbf{L_{disk}(r)}|$ is the unit vector in the direction of disk's angular momentum at radius r. In the presence of small warp ($\hat{\psi} > 0$, but $\ll 1$), strong oscillatory motions are generated in a tilted disk due to warp-induced radial pressure gradients \citep{Ogilive_2013}. However, when $\hat{\psi}(r) > 1$, nonlinear features can emerge. This occurs because, in this regime, local warp surpasses the disk's height, causing these oscillatory motions to couple with the warp. As a result, the radial velocity approaches the sound speed \citep{Sorathia_2013,Sorathia_2014}, leading to compression.

Figure \ref{fig:warp} shows that $\hat{\psi}(r) > 1$ occurs in the inner flows for spin $a = 0.75$ and $0.94$ (shaded regions). This aligns neatly with the region of partial-alignment and coincides with the area where shocks form, as observed in the Figure \ref{fig:general_characteristics}. Notably, we do not identify any region with $\hat{\psi}(r) > 1$ for the $a = 0.5$ case, as the warping in the lowest spin scenario is less pronounced due to minimal twisting and bending of the disk. In this case where $\hat{\psi}(r) < 1$ we do not presume the absence of non-linear effects, but rather that they are much weaker than in the higher spin cases. It is essential to note that the condition $\hat{\psi}(r) > 1$ is based on simple physical arguments and should not be interpreted rigidly, i.e. \cite{Sorathia_2014} chose an upper limit of 0.8 instead of 1. 

\begin{figure}
\centering
\begin{minipage}{.5\textwidth}
  \centering
  %\hspace*{0cm}
  \includegraphics[width=0.75\linewidth]{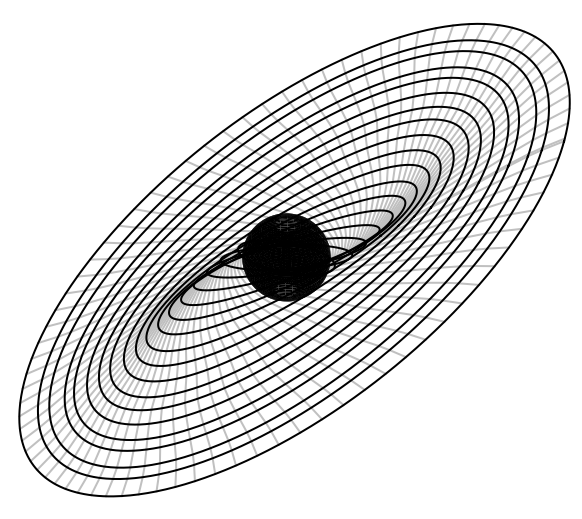}
  \caption{Representation of the averaged disk plane, depicted as tilted and twisted rings for the $a \approxeq 0.94$ case. The visualization emphasizes pronounced warping, particularly near the black hole (black sphere). The BH spin axis is vertical in the figure, where the viewpoint camera is located at an elevation of -$130^\circ$ and an azimuth of -$90^\circ$ with respect to the BH equatorial plane.  }
  \label{fig:Ogilive's warped disk}
\end{minipage}
\end{figure}

Figure \ref{fig:Ogilive's warped disk} visualizes the averaged disk plane, represented as a series of rings observed from the perspective of the BH frame. Each ring is characterized by its individual tilt and twist, in accordance with the radial profiles displayed in Figure \ref{fig:inclination_and_twist}. Notably, the figure distinctly showcases the pronounced warping of the disk, particularly evident in the vicinity of the black hole where the twisting is more intense.

\subsection{Origin of the shocks}\label{sec:shockformation}
\begin{figure*}[t!]
\centering
\includegraphics[width=\textwidth]{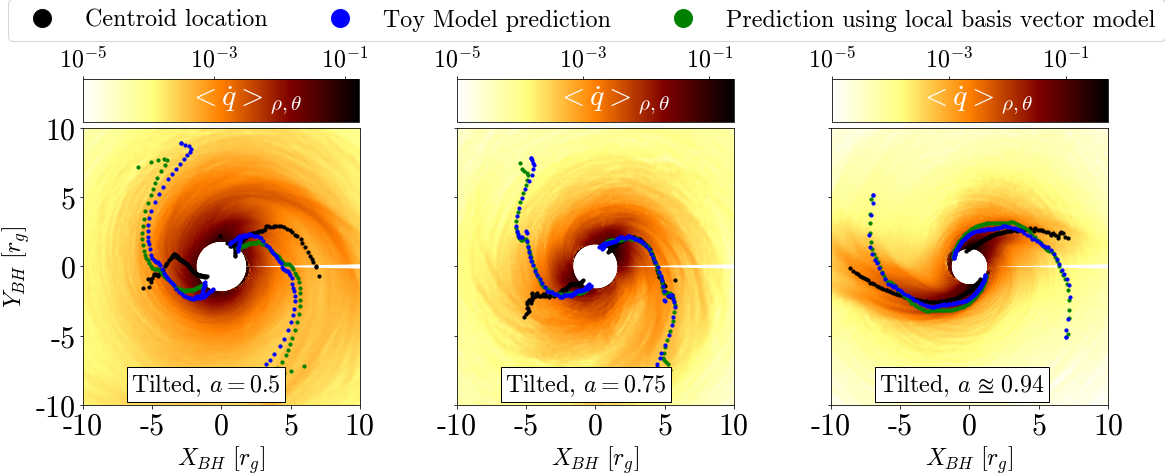}
\caption{\small{Density-weighted, vertically integrated snapshots of the heating rate ($\dot{q}$) with centroid positions (black) for all tilted flows, overlaid with locations of maximum crowding of orbits from the toy model (blue) and the predicted shock locations from the basis vector model (green). The close agreement between the centroid locations and predicted shocks location from both methods at BH spins, $a = 0.75$ and $\approxeq 0.94$, suggests that the shocks form due to the crowing of elliptical orbits in the flow. The significant discrepancy between the black and blue/green points for radii $ r > 6~r_g$ is attributed to the large radial variation in the tilt profile compared to the twist.}}
\label{fig:toy_model}
\end{figure*}

To understand the formation of shocks and how they are affected by radial variations of tilt, and twist, we employed two different approaches. First, we draw insights from the work of \cite{Ivanov_Illarionov_1996} and developed a toy model following the prescription below:
\begin{enumerate}
    \item We define a series of centered elliptical orbits with an eccentricity of $e = 0.1$\footnote{Based on our simulations, we observe that disk eccentricities typically range between $0.01$ and $0.15$. Given that radial motions flip sign above and below the local disk midplane \cite{Fragile_2008,Kaaz_2023}, we believe the averaged fluid orbits are only slightly perturbed circular orbits with a small eccentricity.} in a disk-aligned frame, with the BH at the center. The disk-aligned frame\footnote{The relationship between the angular coordinates of our simulation setup and the disk-aligned frame (or tilted frame) is given in Appendix \ref{sec:Appendix_B}.} is oriented such that its z-axis ($\equiv Z^\prime$) is parallel to the disk's angular momentum. The x-axis ($\equiv X^\prime$) is aligned with the line of nodes, and the y-axis ($\equiv Y^\prime$) completes the right-handed coordinate system.  Initially, these orbits are untwisted, centered and are constrained to the $X^\prime-Y^\prime$ plane.
    \item We then represent these orbits in the BH reference frame by calculating the angular coordinates $(\theta,\phi)$ using Equation \ref{eq:tilted_frame}. This process manifest the geometry of the averaged tilted disk as seen from the BH frame. Figure \ref{fig:Ogilive's warped disk} shows the orientation of such orbits viewing at a specific angle for the $a \approxeq 0.94$ case.
    \item To locate the regions of maximum crowding, we identify the points where the distance between neighboring rings is minimized. These points are labeled as the sites of maximum crowding, which we predict to be the locations where shocks form.
    \item Finally, we superimpose the predicted sites of maximum crowding on the plot of vertically-integrated heating rate $\left(\langle\dot{q}\rangle_{\rho,\theta}\right)$ and compare them with observed shock-forming regions.
\end{enumerate}

Second, based on the work of \cite{Ogilive1999,Ogilive_2013}, we defined local basis vectors for our warped disk with a few adjustments:
\begin{enumerate}
    \item $\mathbf{m}$ = $\frac{r d \mathbf{l(r)}}{d r}$, a unit vector orthogonal to $\mathbf{l}$ and pointing where adjacent warped annuli are most separated.
    \item $\mathbf{\hat{n}}$ = $\frac{\mathbf{l} \times \mathbf{m}}{\left| \mathbf{l} \times \mathbf{m}\right|}$, a unit vector pointing where adjacent warped annuli are least separated, and thus should point to the location of the shocks.
    \item Given the $m = 2$ nature of shocks, we also considered the vector $- \mathbf{\hat{n}}$, as both $\mathbf{\hat{n}}$ and $-\mathbf{\hat{n}}$ should point towards the least separated annuli.
    \item Finally, we superimposed the $x$ and $y$ components of both vectors, $r \mathbf{\hat{n}}$ and $-r \mathbf{\hat{n}}$ on the $\left(\langle\dot{q}\rangle_{\rho,\theta}\right)$ plot, where we multiply by $``r"$ to account for the shocks differing locations at different radii.
\end{enumerate}

The findings from both our toy model and basis vector model are illustrated in Figure \ref{fig:toy_model}, where the black dots represent the centroid of vertically-averaged heating rate, and blue and green dots represent the predicted sites of shock location using the toy model and basis vector model, respectively. Both methods predict the shock locations quite well, especially for higher BH spins ($a \approxeq 0.94$ and $0.75$). However, mapping the shock structure in the heating rate plot for the low spin case requires a comprehensive fluid trajectory analysis, as done by \cite{Generozov_2013}, for a meaningful comparison. Because shocks are weaker at lower spins, their $m =2$ structure is less distinct. Consequently, using the centroid positions of the vertically-integrated heating rate as a proxy for shock locations introduces errors, particularly at larger radii and for the lowest spin.

\begin{figure*}[t!]
\centering
\includegraphics[width=0.5\textwidth]{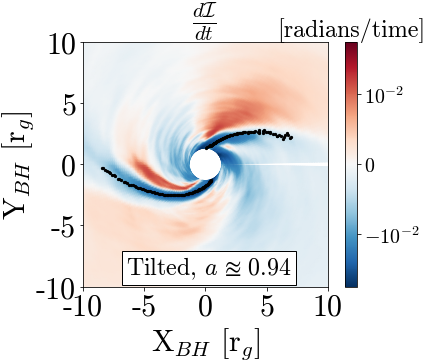}
\caption{\small{Density-weighted, vertically integrated snapshots of change in the rate of inclination for BH spin, $a \approxeq 0.94$. The sound agreement between the shock-forming locations depicted by centroid of $\left(\langle\dot{q}\rangle_{\rho,\theta}\right)$ (black) and the regions where the tilt is decreasing demonstrates the possible role of shocks in partially aligning the inner disk with the BH. }}
\label{fig:didt_highspin}
\end{figure*}

Since both methods are majorly governed by tilt and twist values, and their radial variation, we found that in the very vicinity of the BH, $(r \leq 6~r_g)$,  twist is the dominant factor in determining the spatial structure of the shocks (see Appendix \ref{sec:Appendix_E}) Whereas at larger radii $(r > 6~r_g)$, the observed disparities between our sites of maximum crowding and the centroid positions are primarily due to the radial dependence of the tilt profile, as at these radii the twist in the disk changes more gradually with radius compared to the tilt. Thus, for radii $ r > 6~r_g$ , the predicted shock locations are significantly influenced by the radial variation of the tilt profile. Our observations also indicate that while eccentric orbits are essential in our toy model for identifying regions of local density enhancement, the geometry of these shocks remains largely unaffected as long as the eccentricity remains small $(e < 0.2)$.

\subsection{Shock-induced partial alignment}
\label{sec:timescales}
To understand the decrease in inclination within the inner flow, we investigate the rate of change of inclination along the path of a fluid parcel. This involves calculating the material derivative of inclination ($\mathcal{I}$) using:
\begin{equation}
\label{eq:material_derivative}
    \frac{d \mathcal{I}}{dt} = \frac{\partial \mathcal{I}}{\partial t} + \mathbf{V}\cdot\mathbf{\nabla}{\mathcal{I}}
\end{equation}
where $\textbf{V}$ is the coordinate three-velocity, the components of which are given by: $V^i = u^i/u^t$. 

Given our focus on the steady-state behavior of tilted flows, the first term on the right-hand side of the equation becomes zero. Expanding the second term yields three additional terms that describe the change in inclination of a fluid parcel subjected to a space-dependent velocity field. To assess the influence of shocks, we compute the vertically-integrated rate of change of inclination, represented as:
\begin{equation}
\label{eq:RHS_breakdown}
    \langle \mathbf{V}\cdot\mathbf{\nabla}{\mathcal{I}}\rangle_{\rho,\theta} = \langle V^r \rangle_{\rho,\theta}\frac{\partial \mathcal{I}}{\partial r} + \langle V^\phi \rangle_{\rho,\theta}\frac{\partial \mathcal{I}}{\partial \phi}
\end{equation}
All the quantities in this equation are density-weighted and vertically averaged within the $\sigma \leq 1$ region to only consider the dense area of the flow and reduce errors associated with high radial velocities in polar regions. In the above equation $\mathcal{I}$ is calculated using: 
\begin{equation}
\label{eq:Inclination_2D}
    \mathcal{I}(r,\phi) = \cos^{-1}\left(\frac{\langle L_z \rangle_{\rho,\theta}}{\sqrt{\langle L_x \rangle_{\rho,\theta}^2 + \langle L_y \rangle_{\rho,\theta}^2 + \langle L_z \rangle_{\rho,\theta}^2}}\right)
\end{equation}

Figure \ref{fig:didt_highspin} illustrates the evolution of inclination for a fluid parcel moving in the Eulerian frame for the high-spin case ($a \approxeq 0.94$). We chose the high spin case because the shock structure is well-defined, and the decline in inclination is much more pronounced compared to lower spins. The plot shows that the region with a negative rate overlaps with the shock-forming region, identified by black dots from the centroid positions of vertically-averaged heating rate. Additionally, the rate exhibits a non-axisymmetric structure similar to the heating rate [cf. Figure \ref{fig:general_characteristics}]. This indicates that the tilt of a fluid parcel decreases as it passes through a shock front, demonstrating the critical role of shocks in reducing the tilt of inner flows. However, fluid parcel also experiences an increase in tilt as it moves out of shock forming regions. This feature, where $\frac{d \mathcal{I}}{dt}$ is positive, can be attributed to the angular momentum fluxes along with the gravitational torque.

However, it is important to note that the centroid positions do not mark the exact location of the shocks front. Moreover, calculating the rate of change of inclination in the Eulerian frame involves numerical derivatives, which can introduce small errors. Despite these limitations, the regions with decreasing inclination are consistent with the shock-forming regions. The rate of change of inclination for $a = 0.5$ and $0.75$ are shown in Figure \ref{fig:Appendix:didt_otherspins}.

Furthermore, we observe that the second term on the right-hand side of equation \ref{eq:RHS_breakdown}\textemdash which represents the flux transport of the change in inclination of a gas parcel through azimuthal motion, is dominant and largely determines the $\mathbf{V}\cdot\mathbf{\nabla}{\mathcal{I}}$. This causes the sign of $d\mathcal{I}/dt$ to flip at the shock position. To ensure the robustness of this result, we performed additional analyses where we defined the rate of change of inclination in three different ways, as detailed in Appendix \ref{sec:Appendix_D}. Although these definitions showed minor differences, the overall features of the rate of change of inclination variable remained consistent.

\begin{figure}[t!]
  \centering
  \hspace*{0cm}
  \includegraphics[width=1.\linewidth]{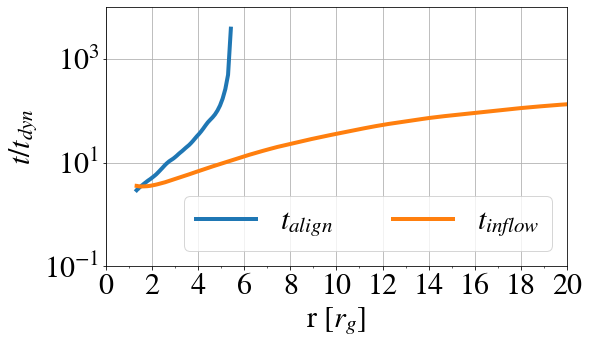}
  \caption{\small{Plot comparing the alignment timescale (blue) to the accretion timescale (orange) within the disk. Both timescales are normalized by the local orbital period ($t_{dyn}$). The fact that the gas is accreting much faster than it is aligning indicates the inability of the inner disk to fully align with the spin of the BH.  }}
  \label{fig:timescales}
\end{figure}

A natural question arises: if shocks are promoting alignment within the inner disk, why do we not observe complete, mutual alignment between the disk and the black hole, akin to the Bardeen-Petterson effect? Why is partial alignment the prevailing outcome? To address this question, we introduce an alignment timescale: $t_{align} \equiv \mathcal{I}(r,t)/\left(-d \mathcal{I}(r,t)/d t\right)$. The inclusion of a minus sign ensures a positive timescale, considering our expectation that $d \mathcal{I}(r,t)/d t < 0$ during the alignment process.

We compared this timescale with the inflow timescale: $t_{inflow} = r/(-\langle V^r \rangle_{\rho})$. Our hypothesis is that the disk does not completely align because the gas is forced to flow radially inward at a much faster rate than the time it requires to dissipate its out-of-plane angular momentum entirely. Thus, we examine the relationship between the alignment timescale and the inflow/accretion timescale. In the above definitions, $~\langle V^r \rangle_{\rho}$ represents density-weighted, time averaged radial velocity. The alignment timescale is calculated by taking the ratio of radial inclination profile (cf. Figure \ref{fig:inclination_and_twist}) to the azimuthal average of $d \mathcal{I}/d t$ that we calculated using Equation (\ref{eq:material_derivative}).

Figure \ref{fig:timescales} illustrates these timescales as a function of radius for the misaligned flow around the BH of spin, $a \approxeq 0.94$. The y-axis is normalized by the dynamical timescale ($t_{dyn} = 1/\Omega$, where $\Omega = \frac{d \phi}{d t} = \langle V^\phi \rangle_{\rho}$), representing time in periods of flow rotation. The figure depicts that in the inner regions where the tilt is decreasing, $t_{align}$ exceeds $t_{inflow}$ by a factor of $\sim 2 - 100$. This indicates that gas lacks the required time to attain mutual alignment at a specific radius since it's compelled to migrate inward at a considerably faster pace, preventing it from fully shedding its out-of-plane angular momentum, and hence, leaving the inner disk to partial-alignment only.

\section{Conclusions \& Discussion}
\label{sec:conclusion}
We have presented a series of global, idealized 3D GRMHD simulations of misaligned accretion onto a black hole with different spins, ranging from $a = 0.5$ to $0.9375$, with an initial inclination of $24^\circ$. Comparing the misaligned simulations to an aligned accretion flow reveals that in all tilted flows, the disks evolve to a highly warped and twisted steady-state structure showing no signs of the Bardeen-Petterson alignment \citep{Bardeen_Petterson}. We do, however, discover a potential aligning process for weakly-magnetized, thick accretion disks ($H/r > \alpha$) that brings the inner disk into partial alignment with the BH. The standing shocks in the tilted flows are torquing the inner disk into the BH equatorial plane, likely through the redistribution of angular momentum via the redirection of fluid elements. For instance, while analyzing the evolution of inclination of a comoving fluid [cf. Equations (\ref{eq:material_derivative}) and (\ref{eq:RHS_breakdown})], we observe that the decrease in inclination at the shock location is far greater due to azimuthal motion than radial motion, suggesting the orbital mixing of fluid elements of different orientations of angular momentum. This importance of azimuthal mixing is also hinted at \cite{Liska2018} and \cite{Sorathia_2013}, as \cite{Sorathia_2013} postulated that for alignment to occur azimuthal mixing of angular momentum is required\textemdash the second term in equation (\ref{eq:RHS_breakdown}) may quantify such a mechanism. Overall, the presence of spiral shocks results in an even angular momentum distribution (or, less non-axisymmetric) in the azimuthal direction. 

However, it is also possible that the partial alignment we observe is a result of the oscillatory nature of bending waves, as suggested by \cite{Ivanov_Illarionov_1996} and \cite{Zhuravlev_2014}. Given that the nonlinear features arise close to the BH, this decrease in inclination could result from oscillatory behavior associated with standing, nonlinear bending waves.

Our study reveals that the strength of shocks increases with the BH spin, in line with findings of \cite{White_2019}. This relationship offers an explanation for the absence of shocks observed by \cite{Teixeira_2014} for a BH spin of $a = 0.1$. Our findings that shocks partially align the inner flow is consistent with the steeper alignment rate observed in high-spin cases. \cite{White_2019} demonstrated that shocks become more pronounced as the initial misalignment of the disk increases for the same BH spin. Combining this finding with our shock-induced partial alignment, it is conceivable that in severely misaligned flows, the inner flow may undergo complete alignment with the spin of the BH.  In future research, we intend to develop models to quantify the strength of shocks as a function of varying initial tilt and BH spins, enabling us to better understand the process of alignment between a disk and its black hole.

While our simulations exhibit general characteristics consistent with previous GRMHD studies of hot, tilted flows \citep{Fragile_2007,Fragile_2008,Generozov_2013,Teixeira_2014}, and to a certain extent in warped thin disks \citep{Kaaz_2023}, we did not observe definitive evidence of significant global precession in our system. In the case of high BH spin ($a \approxeq 0.94$), simulated over an extensive period of $t \approx 25,000r_g/c$, our warped disk exhibited only a modest precession of $\approx 10^\circ$ for the radial shell of $15$ to $50~r_g$, suggesting a global precession timescale of $\approx 10^6 r_g/c \approx 0.5(M/M_\odot)s$. This timescale is consistent with the predictions derived from the formulation provided by \cite[Equation (43)]{Fragile_2007}, where we took inner and outer radius of the evolved disk to be, $r_i \sim 15~r_g$, $r_0 \sim 50~r_g$ as fiducial parameters\footnote{We acknowledge the fact that inner and outer radius are not well constrained parameters; nevertheless, our fitted precession timescale remains of similar order, with $r_{in} \approx 6 ~r_g$ and $r_{out} \approx 100~r_g$. } , and $\zeta \sim -0.83$ (derived from $\Sigma = \Sigma_i (r/r_i)^{-\zeta}$) as estimated in our study. This observation suggests the possibility of the tilted disk precessing as a rigid body under the influence of Lense-Thirring torque \citep{Lense-Thirring}. For lower spins, however, the simulation duration was insufficient to capture significant precession, thus limiting our ability to draw meaningful conclusions.

The results of our toy model confirm the arguments of \cite{Ivanov_Illarionov_1996} and their close agreement with the local basis vector method suggests that nodes of the shocks form predominantly at $\pm r \mathbf{\hat{n}}$, i.e. at the location where the adjacent warped annuli are least separated. Additionally, the simplicity of our model enables us to quantify the effects of tilt and twist. where we discover that the twist predominantly governs the geometry of the shocks, but their location can change if the disk is significantly misaligned. Moreover, while global precession was not prominently observed in our simulations, it is conceivable that, akin to the findings of \cite{Fragile_2008}, spiral shocks might have precessed on a similar timescale if global precession had been present. Considering the pivotal role of twist in determining the locations of maximum crowding, it is plausible that our toy model could predict the location of shocks in scenarios involving global precession of the disk. 

Our model does, however, have limitations. First, it cannot predict the geometry of the shock from the initial misalignment of the torus. Complete knowledge of tilt and twist is needed to predict the spatial structure of shocks. One can avoid performing the simulations by predicting the tilt and twist profile using the linear theory, but as shown by \cite{White_2019}, the problem is highly non-linear. Secondly, because we use the radial profiles of tilt and twist, we could not predict the vertical structure of the shocks. To determine this we would need to completely track the full parameter variations, such as density, pressure, and velocity across the shock front. Despite having these limitations, our model shows great potential for locating shocks and can be applied to observations of warped disks. While it is true that finding the tilt and especially twist of real warped disks is a highly complex task, one could take the jet inclination as an average tilt and use this as an initial condition in simulations to potentially predict where shocks will develop. 

The increase in inclination at the outer radii is consistent with the physics of bending waves. However, as shown by \cite{White_2019}, the straightforward application of the linear theory does not properly yield the observed profile of tilt. The high sensitivity of linear theory to the surface density profiles, Shakura-Sunyaev $\alpha$ viscosity, and the number of assumptions that are violated in GRMHD simulations make the predictions of linear theory inapplicable. An uncharted territory being explored by \cite{Zhuravlev2011} is deriving the governing equations for each component of angular momentum and assessing the contribution of the torque produced due to frame-dragging of the spacetime, pressure forces, Reynolds stress, and Maxwell stress. Approaching the problem in the said manner may also help in determining the source of shocks-induced partial alignment of the inner disk and the dominance of $V^\phi\frac{\partial \mathcal{I}}{\partial \phi}$ in governing $d\mathcal{I}/dt$.  

The timescale analysis shows why the inner disk never fully aligns. We define an alignment timescale and observed that $t_{align} > t_{inflow}$, suggesting that the gas does not have enough time to dissipate its out-of-plane angular momentum. While this is a simple, dimensionality-based estimate, it explains our results quite well and we believe it to be a good order of magnitude check. Also, as one would expect from the ``alignment timescale'', our definition is moot in the outer regions of the disk where the flow is significantly more misaligned. 

Finally, standing shocks can accelerate the plasma, and can be the sites of particle acceleration and nonthermal electron distribution in collisionless accretion flows like for Sgr A$^*$ \citep{Fragile_2008}. We suggest that a radiative 3D GRMHD simulation of a misaligned accretion disk rotating around a fast-spinning BH can help us to identify the signatures of standing shocks in spectra. It may also help us to quantify the difference between the spectrum of aligned and misaligned flows. Furthermore, it will allow us to investigate the arguments of \cite{Zhuravlev2011}, who claimed that strong irradiation of the outer parts of the disk by the inner parts could change the spectra of the disk. To effectively probe the effects of shocks, we also propose a parameter survey similar to this, incorporating passive scalars as tracers to investigate the fluid trajectories across the shocks front.

\section{Acknowledgments}
We thank M. C. Begelman, C. J. White, K. Long, C. Echibur\'u -Trujillo, and N. Scepi for stimulating discussions related to this work. We thank the anonymous reviewers for their insightful comments and suggestions, which have significantly improved our research. This work was supported in part by National Science Foundation award AST-2034306, the NASA Astrophysics Theory Program grant 80NSSC20K0527, and by an Alfred P. Sloan Research Fellowship (JD).

\vspace{5mm}

\software{Matplotlib \citep{Matplotlib}, NumPy \citep{NumPy}}

\appendix

\section{\texorpdfstring{Properties of $\MakeLowercase{a}= 0.5$ \MakeLowercase{and} $0.75$ tilted flows}{Properties of a = 0.5 and 0.75 tilted flows}}
\label{sec:Appendix_A}
\begin{figure*}[t!]
\centering
\includegraphics[width=\textwidth]{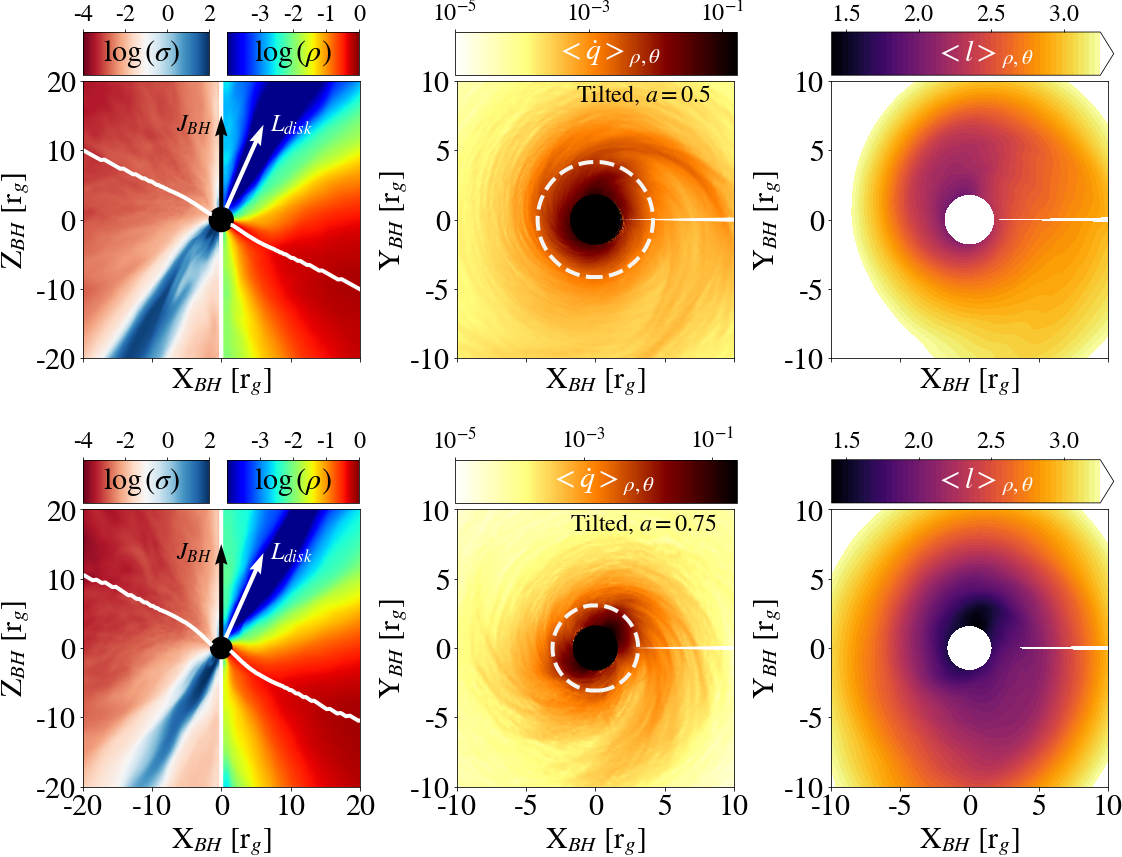}
\caption{\small{Similar to Figure \ref{fig:general_characteristics}, the above plot shows the properties of the tilted flows for BH spin, $a = 0.5$ and $0.75$. Warping of the disk and standing shocks are apparent from poloidal slices of $\rho -\sigma$ and density-weighted, vertically averaged $\langle\dot{q}\rangle_{\rho,\theta}$ and $\langle l \rangle_{\rho,\theta}$ profiles. The strength of the shocks is shown to be proportional to the BH spin. However, unlike shocks strength, the strongly magnetized region shrinks with increasing spin due to strong vertical motions that result from enhanced warping.}}
\label{fig:Appendix:general_characteristics}
\end{figure*}

The purpose of this appendix is to describe the properties of tilted flows for spins $a = 0.5$ and $0.75$. In the section \ref{sec:generalcharacteristics}, we describe the warping of the disk and the emergence of non-axisymmetric features in vertically-integrated profiles of $\langle\dot{q}\rangle{\rho,\theta}$ and $\langle l \rangle{\rho,\theta}$  for spin $a \approxeq 0.94$. These profiles reveal the presence of standing shocks. Similar characteristics are observed in tilted flows with spins $a = 0.5$ and $0.75$, as depicted in Figure \ref{fig:Appendix:general_characteristics}.

The warping of the disk is evident from poloidal slices of density-magnetization profiles, $\rho-\sigma$. As expected, the low-density funnel region, aligned with areas of high magnetization ($\sigma > 1$), stands perpendicular to the disk midplane. Comparing $\sigma$ profiles of tilted flows with spins $a = 0.5$, $0.75$, and $\approxeq 0.94$ (Figure \ref{fig:general_characteristics}), we note a reduction in the regions with strong magnetization as the BH's spin increases. This phenomenon is attributed to more substantial vertical motions, notably noticeable in the high-spin case, demonstrated in Figure \ref{fig:inclination_and_twist}, which exhibits pronounced warping. 

The vertically averaged snapshots of the heating rate ($\dot{q}$) reveal the presence of standing shocks.  Notably, the degree of non-axisymmetry intensifies as the BH's spin increases, suggesting that the strength of the shocks and their impact on disk partial-alignment and angular momentum transport escalates with higher BH spin.
\section{Tilted frame angular coordinates}
\label{sec:Appendix_B}

To assess parameters within the tilted frame, which aligns with the disk, we utilize angular coordinates ($\theta^\prime$,$\phi^\prime$) linked to grid coordinates ($\theta$,$\phi$) through Equation (7) in \cite{White_2019}. For thoroughness, we present the relationship here: 
\begin{align}\label{eq:tilted_frame}
    \begin{split}
    \theta^\prime &= \cos^{-1}(\cos{\mathcal{I}}\cos{\theta} + \sin{\mathcal{I}}\cos{\mathcal{T}}\sin{\theta}\cos{\phi} + \sin{\mathcal{I}}\sin{\mathcal{T}}\sin{\theta}\sin{\phi})~, \\
    \phi^\prime &= \tan^{-1}(-\sin{\mathcal{T}}\sin{\theta}\cos{\phi} + \cos{\mathcal{T}}\sin{\theta}\sin{\phi}, -\sin{\mathcal{I}}\cos{\theta} + \cos{\mathcal{I}}\cos{\mathcal{T}}\sin{\theta}\cos{\phi} + \cos{\mathcal{I}}\sin{\mathcal{T}}\sin{\theta}\sin{\phi})~, \\
    \theta &= \cos^{-1}(-\sin{\mathcal{I}}\sin{\theta^\prime}\cos{\phi^\prime} +
    \cos{\mathcal{I}}\cos{\theta^\prime})~, \\
    \phi &= \tan^{-1}(\sin{\mathcal{I}}\sin{\mathcal{T}}\cos{\theta^\prime} + \cos{\mathcal{I}}\sin{\mathcal{T}}\sin{\theta^\prime}\cos{\phi^\prime} + \cos{\mathcal{T}}\sin{\theta^\prime}\sin{\phi^\prime}, \\
         &\qquad \qquad \sin{\mathcal{I}}\cos{\mathcal{T}}\cos{\theta^\prime} + \cos{\mathcal{I}}\cos{\mathcal{T}}\sin{\theta^\prime}\cos{\phi^\prime} - \sin{\mathcal{T}}\sin{\theta^\prime}\sin{\phi^\prime})
    \end{split}
\end{align}

\section{\texorpdfstring{Rate of change of inclination for $\MakeLowercase{a}= 0.5$ \MakeLowercase{and} $0.75$ tilted flows}{Rate of change of inclination for a = 0.5 and 0.75 tilted flows}}

\label{sec:Appendix_C}
\begin{figure*}[t!]
\centering
\includegraphics[width=\textwidth]{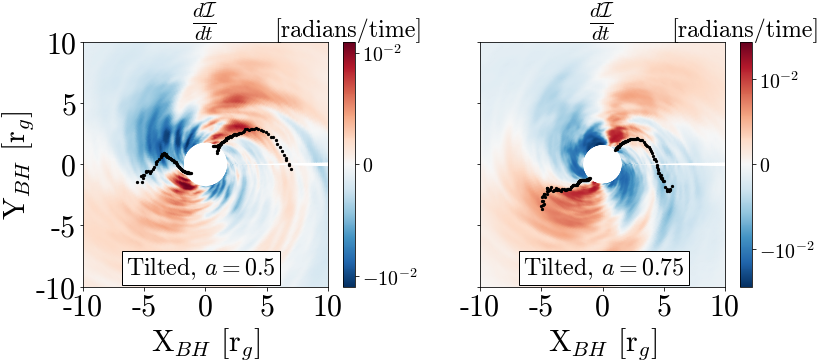}
\caption{\small{Similar to Figure \ref{fig:didt_highspin}, the above plot shows the change in rate of inclination for the spins of BH $a = 0.5$ (left) and $0.75$ (right). The good agreement between the black points describing the location of shocks and the regions of negative rate for $a  = 0.75$ case elucidates the impact of shocks in aligning the inner disk. However, because the shocks are weak in the low spin case, the decline we observe in the tilt of the inner disk may be due to numerical errors.}}
\label{fig:Appendix:didt_otherspins}
\end{figure*}
This section delves into the rate of change of inclination for BH spin values of $a = 0.5$ and $0.75$. Figure \ref{fig:Appendix:didt_otherspins} shows that the rate is largely negative in the inner flow, especially for $a = 0.75$. This is consistent with the radial drop in the tilt profile seen in Figure \ref{fig:inclination_and_twist}. The scattered black points in the plot correspond to centroid positions of vertically-integrated heating rate, which we used as proxy for the determining the location of the shocks front. For the $a = 0.75$ scenario, we observe a sound agreement between the positions of shocks and regions exhibiting negative inclination rates. This suggests that the shocks affect the inclination of disk, particularly compelling the inner flow to align with the spin of the BH.

However, for the $a = 0.5$ scenario, the situation is less straightforward. This is because the shocks are relatively weak, suggesting that their influence is subtler and less discernible through our simple analysis of $d\mathcal{I}/dt$. Additionally, the observed tilt reduction for the low spin is less than 2.5$^\circ$, possibly due to errors related to angular momentum projections.

\section{Different methods to compute rate of change of inclination}
\label{sec:Appendix_D}
In this section, we evaluate whether the rate of change of inclination measured in the Eulerian frame at steady-state varies depending on the definition employed. The following table outlines three distinct methods, all based on the same fundamental principle, $\frac{d \mathcal{I}}{dt} = \mathbf{V}\cdot\mathbf{\nabla}{\mathcal{I}}$.

\begin{table}[ht]
\centering
\begin{tabular}{|l|l|l|}
%\begin{tabular}{|p{2.5cm}|p{9cm}|p{4.5cm}|}
\hline
Method & $\langle \mathbf{V}\cdot\mathbf{\nabla}{\mathcal{I}}\rangle$ term & $\mathcal{I}$ term \\
\hline
Method 1 &

$\langle \mathbf{V}\cdot\mathbf{\nabla}{\mathcal{I}}\rangle = \langle V^r \rangle_{\rho,\theta}\frac{\partial \mathcal{I}(r,\phi)}{\partial r} + \langle V^\phi \rangle_{\rho,\theta}\frac{\partial \mathcal{I}(r,\phi)}{\partial \phi}$
&
$\mathcal{I}(r,\phi) = \cos^{-1}\left(\frac{\langle L_z \rangle_{\rho,\theta}}{\sqrt{\langle L_x \rangle_{\rho,\theta}^2 + \langle L_y \rangle_{\rho,\theta}^2 + \langle L_z \rangle_{\rho,\theta}^2}}\right)$
\\
\hline
Method 2 &
$\begin{aligned}
\langle \mathbf{V}\cdot\mathbf{\nabla}{\mathcal{I}}\rangle &= \langle V^r \rangle_{\rho,\theta}\left\langle\frac{\partial \mathcal{I}(r,\theta,\phi)}{\partial r}\right\rangle_{\rho,\theta} \\
&+ \langle V^\theta \rangle_{\rho,\theta}\left\langle\frac{\partial \mathcal{I}(r,\theta,\phi)}{\partial \theta}\right\rangle_{\rho,\theta} \\ 
&+ \langle V^\phi \rangle_{\rho,\theta} \left\langle\frac{\partial \mathcal{I}(r,\theta,\phi)}{\partial \phi}\right\rangle_{\rho,\theta}
\end{aligned}$
&
$\mathcal{I}(r,\theta,\phi) = \cos^{-1}\left(\frac{L_z}{\sqrt{L_x^2 + L_y^2 + L_z^2}}\right)$

\\
\hline
Method 3 &
$\begin{aligned}
\langle \mathbf{V}\cdot\mathbf{\nabla}{\mathcal{I}}\rangle &= \left\langle V^r \cdot \frac{\partial \mathcal{I}(r,\theta,\phi)}{\partial r}\right\rangle_{\rho,\theta} \\
&+ \left\langle V^\theta \cdot \frac{\partial \mathcal{I}(r,\theta,\phi)}{\partial \theta}\right\rangle_{\rho,\theta} \\
&+ \left\langle V^\phi \cdot \frac{\partial \mathcal{I}(r,\theta,\phi)}{\partial \phi}\right\rangle_{\rho,\theta}
\end{aligned}$
&
$\mathcal{I}(r,\theta,\phi) = \cos^{-1}\left(\frac{L_z}{\sqrt{L_x^2 + L_y^2 + L_z^2}}\right)$
\\
\hline
\end{tabular}
\label{Tab:Tcr}
\caption{\small{Comparison of different methods for calculating $\frac{d \mathcal{I}}{dt}$. Method 1 uses vertically-averaged velocities and inclination, Method 2 uses mean fluid velocity and three-dimensional inclination gradients, and Method 3 incorporates both mean and turbulent fluid velocities and three-dimensional inclination gradients.}}
\end{table}

\begin{figure*}[t!]
\centering
\includegraphics[width=\textwidth]{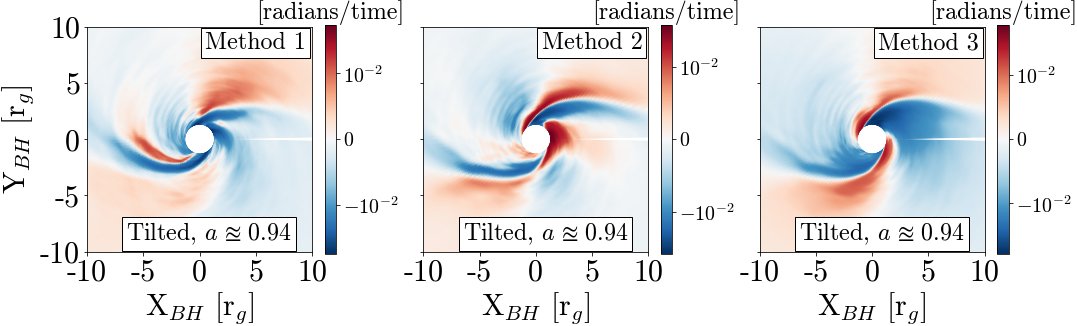}
\caption{\small{The above plot shows the rate of change inclination ($\frac{d \mathcal{I}}{d t}$) for the BH spin $a \approxeq 0.94$, calculated using three different methods. While these methods showed minor differences, the overall features of the plot remained consistent, supporting the reliability of our findings despite the potential numerical errors.}}
\label{fig:Appendix:didt_different_methods}
\end{figure*}

Figure \ref{fig:Appendix:didt_different_methods} presents the resulting calculations using the said methods for the high spin case. As evident from the figure, there are noticeable discrepancies in the measured values of $\frac{d \mathcal{I}}{d t}$, particularly in the vicinity of the BH. The differences between Methods 2 and 3 are straightforward to interpret: Method 2 considers only the change in the tilt of a fluid parcel as it moves with the mean fluid velocity, whereas Method 3 incorporates both mean and turbulent fluid velocities. The deviation between Method 1 and other two methods arises primarily due to the different definitions of inclination we chose.  Despite these variations, the rates measured by all methods display similar non-axisymmetric structures and regions of decreasing inclination, aligning consistently with the non-axisymmetric structure in the vertically-integrated heating rate. We selected Method 1 for our primary analysis because it defines inclination as the angle between the angular momentum vector of the vertically-averaged disk and the BH spin axis.

\section{Geometry of the spiral shock in the disk-aligned frame}
\label{sec:Appendix_E}
To substantiate the robustness of the $m = 2$ standing shocks, we quantified the heating rate within the disk-aligned frame. Initially, we transformed the metric tensor to the tilted frame using the relation $g_{\mu^\prime \nu^\prime} = \frac{\partial x^\alpha}{\partial x^{\mu^\prime}} \frac{\partial x^\zeta}{\partial x^{\nu^\prime}} g_{\alpha \zeta}$, where $g_{\alpha \zeta}$ represents the metric defined in Kerr-Schild coordinates. The Jacobian $(\frac{\partial x^\alpha}{\partial x^{\mu^\prime}})$ for this transformation is detailed in \cite{White_2019}. Subsequently, we computed the area element in the new coordinates as $\sqrt{-g^\prime}$, and calculated the angular coordinates $\theta^\prime$ and $\phi^\prime$ following the equations outlined in Appendix \ref{sec:Appendix_B}. These calculations revealed that at specific radii, the angular coordinates were neither uniformly distributed nor covered the complete polar $([0,\pi])$ and azimuthal $([0,2\pi])$ ranges. 

\begin{figure*}[!ht]
\centering
\includegraphics[width=0.5\textwidth]{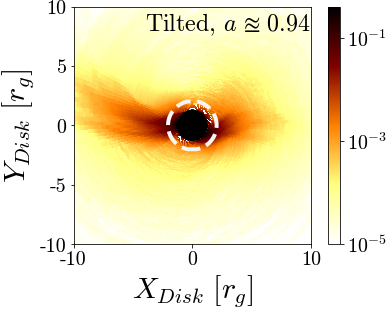}
\caption{\small{The above plot shows the density-weighted, vertically averaged heating rate ($<\dot{q}>_{\rho,\theta^\prime}$) in the tilted frame. Standing shocks predominantly form near $\phi^\prime = 0$ and $\pi$, highlighting the influence of twist in determining the shock locations in the un-tilted BH frame.}}
\label{fig:Appendix:qdotavg_tilted_coord}
\end{figure*}

To address this, we constructed a uniform angular grid and interpolated the relevant physical quantities onto it. The density-weighted vertical integration of the heating rate was then performed using the equation:
\begin{equation}
    \langle \dot{q} \rangle_{\rho,\theta^\prime}(r,\phi^\prime) = \frac{\int\int \dot{q} (r,\theta^\prime,\phi^\prime)~\rho(r,\theta^\prime,\phi^\prime) \sqrt{-g^\prime}d\theta^\prime dt}{\int\int \rho(r,\theta^\prime,\phi^\prime) \sqrt{-g^\prime}d\theta^\prime dt}
\end{equation}
Figure \ref{fig:Appendix:qdotavg_tilted_coord} illustrates these calculations for the high-spin tilted case and corroborates the persistence of the standing shocks. Notably, the nodes of the shocks predominantly align with the $Y_{Disk} = 0$ plane, corresponding to a local precession angle of $\phi^\prime = 0$ and $\pi$. This alignment explains the observed influence of the twist angle on the spatial configuration of the shocks in the original BH frame.

\bibliography{referencesfile}{}
\bibliographystyle{aasjournal}



\end{document}